\documentclass[
  aps,
  PRL,
  reprint,
  superscriptaddress,
  amsmath,amssymb,
  floatfix,
  nofootinbib
]{revtex4-2}
\usepackage{float}
\usepackage{graphicx}
\usepackage{bm}
\usepackage{braket}
\usepackage{booktabs}
\usepackage[colorlinks=true,linkcolor=blue,citecolor=blue,urlcolor=blue]{hyperref}
\graphicspath{{figures_exp01/}{figures/}}
\usepackage{orcidlink}
\begin{document}

\title{Surface-Code Quantum Error Correction for Molecular Tweezer Arrays: Encoding, Layout, and Correlated Noise}
\author{Niharika Verma\,\orcidlink{https://orcid.org/0009-0008-0339-6588}}
\email{niharikachandraverma@gmail.com}
\affiliation{Department of Physics, Indian Institute of Technology Patna, India}

\author{Utpal Roy\,\orcidlink{https://orcid.org/0000-0003-2500-078X}}
\email{uroy@iitp.ac.in}
\affiliation{Department of Physics, Indian Institute of Technology Patna, India}

\begin{abstract}

Polar molecules trapped in optical tweezer arrays offer a promising platform for quantum information processing, providing precise control and long-range interactions that enable high-fidelity gate operations. We investigate quantum error correction in this system and show the influence of underlying physical noise. A mapping is constructed from a molecular tweezer array onto a rotated surface code in which a single specification of the array, namely, which code qubits share a molecule and where those molecules are located, determines both the correlated erasure structure and the dipolar exchange graph. We compare molecular encodings with rotational qudit dimensions ($D$), two and four under heralded molecular loss, coherent dipolar exchange, and imperfect heralding. It is found that a $D=4$ encoding with spatially dispersed pairing exhibits a finite-distance crossing at approximately the same per molecule loss rate as $D=2$, while using $48$-$49\;\%$ fewer molecules, at the cost of a $6$-$10\; \%$ increase in sub-threshold logical error. Fixing the encoding and varying only the spatial embedding produces substantially larger effects: pairing the two co-located qubits along a lattice direction yields a logical-sector asymmetry of approximately fifty times. Over the simulated distances ($d=5,\;7,\;9$), the disfavoured sector shows little or no suppression of logical error with increasing code distance, whereas the favoured sector improves by a factor of $1.5$-$2.3$. We also find that Pauli twirl of the exchange interaction overestimates the logical error rate, which we attribute to the excitation-conserving structure of the interaction. These results reveal that the spatial embedding of correlated loss units is an important design parameter for molecular architectures and that single-sector benchmarks may be insufficient when correlated loss has directional structure.

\end{abstract}

\maketitle

\section{Introduction}

The rotational degrees of freedom of polar molecules provide long-lived internal states, while their dipole-dipole interactions enable strong, tunable couplings between individually trapped molecules. These properties make polar molecules trapped in optical tweezer arrays a promising platform for quantum information processing~\cite{demille2002,Ni_2018,bao2023}. Recent work has further established scalable qudit encodings and entangling gate protocols for molecular tweezer arrays,
including encodings based on rotational manifolds of dimension
$D\!=\!2$-$5$~\cite{muminov2026}. Throughout, we have used $D$ as the dimension of
the molecular rotational manifold used for encoding. In parallel, quantum error correcting encodings
native to individual molecules have been developed by exploiting the structure of
their internal spin and rotational
manifolds~\cite{albert2020,chizzini2022,omanakuttan2023}. These developments
establish molecules as a viable hardware substrate for storing and manipulating
quantum information. 

A complementary question, and the one we take up here, is how the distinctive physical structure of this substrate appears from the perspective of a quantum error correction decoder. Our aim is to characterize how the physical structure of molecular tweezer arrays manifests at the level of surface-code decoding. We construct a physical to code mapping that places the data
qubits of a rotated surface code on a molecular array, characterize the noise channels induced by the molecular architecture, and extract whatever design guidance follows. The observation that organizes the paper is that the molecular encoding
dimension alone does not fully specify the resulting decoding problem. When multiple
code qubits occupy the same molecule, molecular loss produces correlated erasures
whose geometry depends on how those qubits are embedded in the code lattice. The same
local encoding can therefore produce noticeably different logical behaviour under different spatial embedding.

Three features of the molecular platform is highly motivating in the present context, such as, \emph{i}) molecules
interact through always on dipolar exchange, with couplings that decay as $1/r^3$ and
depend on the relative orientation of the molecular separation and dipole moment.
This interaction differs from the predominantly gate mediated interactions considered
in many other qubit platforms, and it raises the question of whether coherent
exchange must be retained explicitly in a decoding model; \emph{ii}) the $D=4$ molecular
encoding considered in Ref.~\cite{muminov2026} places two code qubits within a single
molecule. Loss of that molecule therefore removes two code qubits simultaneously,
producing a correlated erasure whose structure is determined by their positions in
the surface code lattice; \emph{iii}) molecular loss need not always be heralded, since
population can be transferred to internal states that are not resolved by the
measurement process, producing a mixture of decoder visible and silent loss.

Each of these ingredients has precedents in the quantum error correction literature,
though they have generally been studied in other physical settings or in isolation.
Coherent errors in surface codes can behave differently from their Pauli twirled
counterparts, with the direction and magnitude of the discrepancy depending on the
noise model~\cite{bravyi2018,tomita2014,gutierrez2015}. Correlated errors can also
alter surface code performance when their spatial structure is aligned with the code
geometry~\cite{wang2025correlated}. Erasure based error correction, including
decoding with partial or imperfect knowledge of erasure locations, has likewise been
studied extensively in other hardware platforms~\cite{wu2022,sahay2023,kang2023}. Our
contribution is to bring these decoding considerations together for the molecular
tweezer architecture and to determine which features of the molecular encoding and
layout are consequential for logical performance. That synthesis, and not any single
mechanism, is what we offer.

Our analysis yields three main observations. First, a $D=4$ encoding with a spatially
dispersed pairing reaches a finite distance crossing at approximately the same per
molecule loss rate as the $D=2$ encoding, while using approximately $48$-$49$$\%$
fewer molecules for the distances considered. The reduction in hardware comes with a
modest sub threshold increase in logical error, so what we find is a quantitative
trade off and not a universal advantage for either encoding. Second, the spatial
embedding of the $D=4$ encoding can change the behaviour of the code more
substantially than the encoding choice itself. When the two qubits belonging to each
molecule are placed as nearest neighbours along one lattice direction, the resulting
correlated erasures produce a strong asymmetry between the two logical sectors. 
The surface code distance is denoted by $d$. Over the simulated distances $d=5,\;7,\;9$, the disfavoured sector shows little or no suppression of logical error with increasing code distance, whereas the same
embedding improves substantially in the favoured sector. The spatial arrangement of
correlated loss units therefore appears to be a decoder relevant design parameter in
its own right. Third, the coherent dipolar exchange studied here does not produce the
enhancement of logical error that a direct transfer of single qubit coherent error
intuition would suggest. In the regime we simulate, the Pauli twirled model gives a
larger logical error than the coherent evolution, and we associate this with the
excitation conserving structure of the exchange interaction.

A methodological consequence follows from the second observation. Logical performance
cannot in general be characterized by a single logical sector when correlated loss
has directional structure, and we therefore use
\begin{equation}
P_L^{\mathrm{worst}}=\max\!\left(P_L^{X},P_L^{Z}\right)
\label{eq:pworst}
\end{equation}
as the primary figure of merit when comparing molecular encodings and embeddings.
This choice exposes an architectural effect that a single sector benchmark would
hide, since an embedding that appears favourable in one logical sector can
simultaneously degrade the other. The results should be read within the numerical scope of the study. The erasure
analysis uses the finite distances $d=5,\;7,\;9$ and therefore establishes finite distance
crossing behaviour and relative trends, not asymptotic threshold values. The coherent
exchange calculation is evaluated exactly at $d=3$ using state vector evolution, and
we do not extend it to a threshold analysis. We consequently restrict ourselves to
decoder relevant comparisons that remain supported by the simulated regime: the
encoding dependent molecule count trade off, the embedding induced logical anisotropy,
and the distinct behaviour of coherent exchange under Pauli twirling.

\section{Platform and physical to code mapping}
\label{sec:mapping}

\begin{figure*}
\includegraphics[width=0.9 \textwidth]{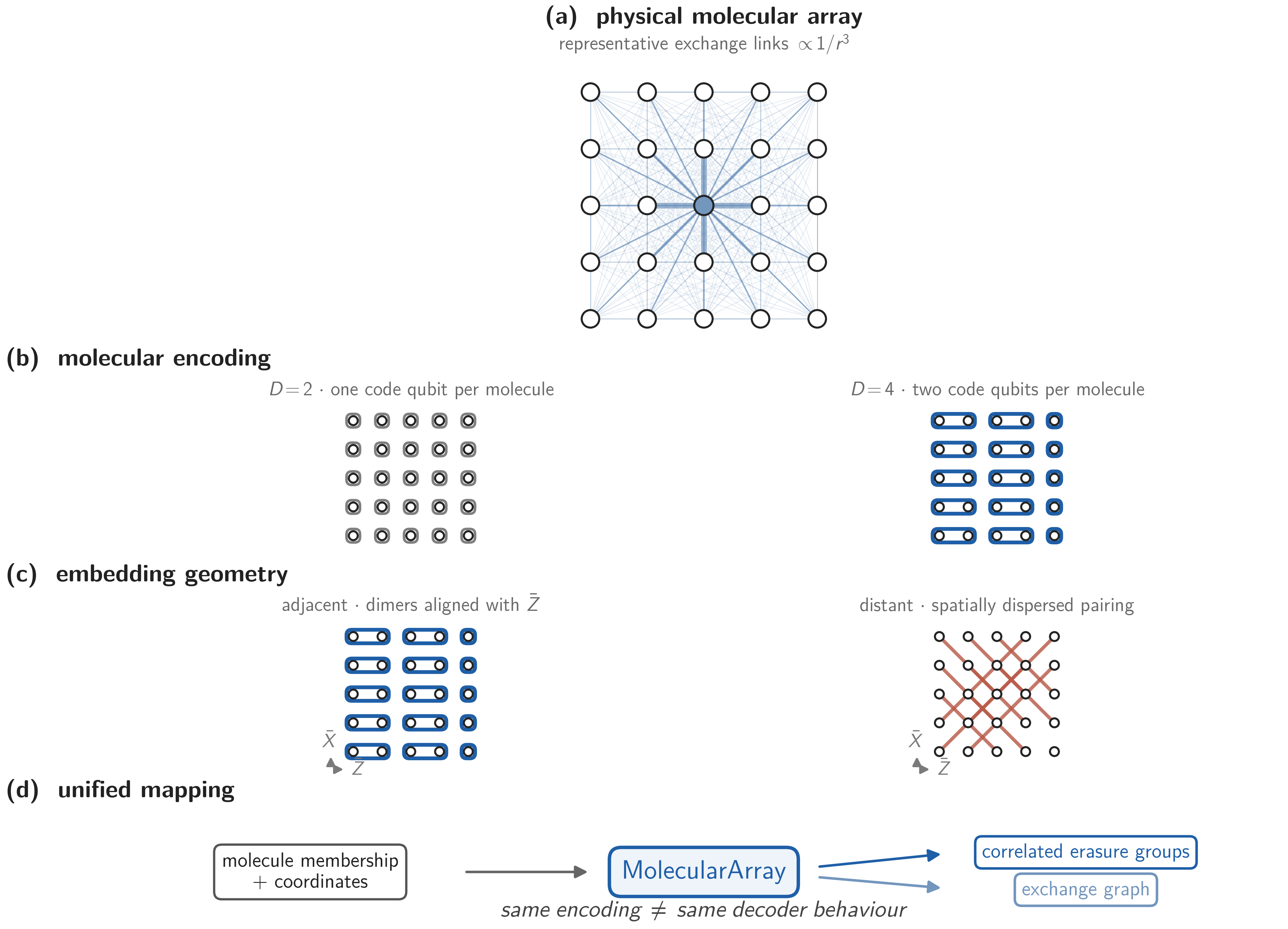}
\caption{Physical to code mapping for a molecular tweezer array hosting the data
qubits of a rotated surface code, illustrated at $d=5$. Circles denote code qubits at
their lattice positions; all panels are generated directly from the layout object used
in the simulations, so the constructions shown are the ones analysed in the text.
(a)~Molecular array with representative dipolar exchange links. The faint mesh marks
nearest neighbour couplings and the heavier links radiate from a single central site,
with line width proportional to $|w_{ij}|$, illustrating the $1/r^{3}$ falloff. Only
these representative links are drawn; the full graph couples every pair of molecular
sites. (b)~The two molecular encodings. For $D=2$, each molecule hosts one code qubit,
so the molecule count equals the qubit count. For $D=4$ each molecule hosts two,
indicated by the capsules, halving the molecule count at the cost of correlated loss.
(c)~The two spatial embeddings of the $D=4$ encoding, with the logical string
directions $\bar{Z}$ and $\bar{X}$ marked by dashed lines. In the adjacent
construction the two qubits of each molecule are lattice neighbours, so every erasure
dimer lies parallel to $\bar{Z}$. In the distant construction the paired qubits are
separated across the lattice and the pairing has no single orientation. The local
encoding is identical in the two cases; only the arrangement of the correlated loss
units differs. Unpaired sites at the right boundary arise from odd $d$ and carry
single qubit erasure. (d)~Both noise structures follow from one specification of the
array. Molecule membership determines which code qubits are erased together when a
molecule is lost, and the molecular coordinates determine the exchange graph. Qubits
sharing a molecule are internal rotational states of the same rotor and are not
mutually dipole coupled, so they contribute to the erasure groups but not to the
exchange graph.}
\label{fig:schematic}
\end{figure*}

We consider $N$ polar molecules held in a two dimensional tweezer array and used to
host the data qubits of a rotated surface code of distance $d$. Two encodings from
Ref.~\cite{muminov2026} are relevant here. In the $D=2$ encoding, a single code qubit
occupies one molecule, so the molecule count equals the qubit count, $N=d^{2}$. In
the $D=4$ encoding, two code qubits share one molecule, and $N\approx d^{2}/2$. Both the noise channels under study are derived from a single specification of the array:
which code qubit belongs to which molecule, and where the molecules sit. We implement this
as one object from which we read off
\begin{noindent}
\begin{enumerate}
\item the \emph{erasure groups}, the sets of code qubits erased together when a
molecule is lost, which follow from molecule membership; and
\item the \emph{exchange graph}, the dipolar couplings
$w_{ij}\propto r_{ij}^{-3}\,(1-3\cos^{2}\theta_{ij})$ between molecular sites,
inherited by the code qubits they host. Qubits sharing a molecule are internal
rotational states of the same rigid rotor and are not mutually dipole coupled, so
intra molecule pairs are excluded from the graph.
\end{enumerate}
\end{noindent}
Figure~\ref{fig:schematic} shows this construction. The point we would draw from it
is that the $D=4$ encoding does not by itself determine the noise the decoder sees.
The same encoding admits different embeddings, and those embeddings produce different
correlated erasure geometries. We consider two. In the \emph{adjacent} embedding the
two qubits of each molecule are nearest neighbours in the code lattice, so every
erasure cluster is a dimer aligned with one lattice direction. In the \emph{distant}
embedding the paired qubits are separated across the lattice, so the clusters are
spatially dispersed. Panel~(c) of Fig.~\ref{fig:schematic} marks the logical string
directions $\bar{Z}$ and $\bar{X}$; in the adjacent case every dimer lies parallel to
$\bar{Z}$.
Molecule counts for the construction are provided in Table~\ref{tab:counts}. The
reduction relative to $D=2$ is close to, but not exactly, a factor of two, and it
differs between embeddings. The adjacent construction pairs qubits row-wise and
leaves one unpaired qubit in each row, giving $45$ molecules at $d=9$, while the
distant construction uses $41$. We therefore quote the actual molecule count for each
embedding instead of assuming an exact factor of two reduction.
\begin{table}
\caption{Molecule counts $N$ for the constructions used here. Unpaired sites arise
from odd $d$ and carry single qubit erasure. The final column gives the reduction
relative to $D=2$ for the distant construction.}
\label{tab:counts}
\begin{ruledtabular}
\begin{tabular}{ccccc}
$d$ & $N_{D=2}$ & $N_{D=4,\mathrm{adj}}$ & $N_{D=4,\mathrm{dist}}$ & reduction \\
\colrule
5 & 25 & 15 & 13 & $48\%$ \\
7 & 49 & 28 & 25 & $49\%$ \\
9 & 81 & 45 & 41 & $49\%$
\end{tabular}
\end{ruledtabular}
\end{table}

\subsection{Noise models and decoding}
We treat the two channels separately, since they act on different timescales and call
for different simulation methods.
\emph{\textbf{Molecular loss}}: In this case, each molecule is lost independently with probability
$p_{\mathrm{mol}}$ per memory experiment, and all code qubits in that molecule then
suffer a random Pauli. A fraction $1-f$ of losses is heralded, and the decoder is
given the locations of the affected qubits as zero weight edges; the remaining
fraction $f$ is silent and receives no herald. Unless stated otherwise $f=0$.
Decoding is minimum weight perfect matching~\cite{pymatching}.

\emph{\textbf{Coherent exchange}}: We evolve the full state vector under
\begin{equation}
H=\sum_{i<j} w_{ij}\,\frac{X_iX_j+Y_iY_j}{2}
\label{eq:hex}
\end{equation}
for an angle $\theta$ per syndrome round, with perfect projective stabiliser
measurement between rounds so that measurement back action is included. The twirled
comparison applies $QUQ$ with a fresh uniformly random $n$ qubit Pauli $Q$ each round,
which realises an exact Pauli twirl of $U=e^{-i\theta H}$ up to a global phase. Both
arms use the same decoder, so any difference between them may be attributed to
coherence alone. This calculation is exact but limited to $d=3$ by the state vector
cost.

\textbf{Figure of merit:} We use $P_L^{\mathrm{worst}}$ of Eq.~\eqref{eq:pworst} for every comparison between
architectures, and report single sector values only where the asymmetry is itself the
object of study. We arrived at this convention through the analysis. An initial single
sector study produced a different architectural ordering, and evaluating both logical
sectors showed that the apparent advantage was a sector dependent effect.

\section{Encoding and embedding trade-offs}
\label{sec:encoding}

\begin{figure*}
\includegraphics[width=0.9\textwidth]{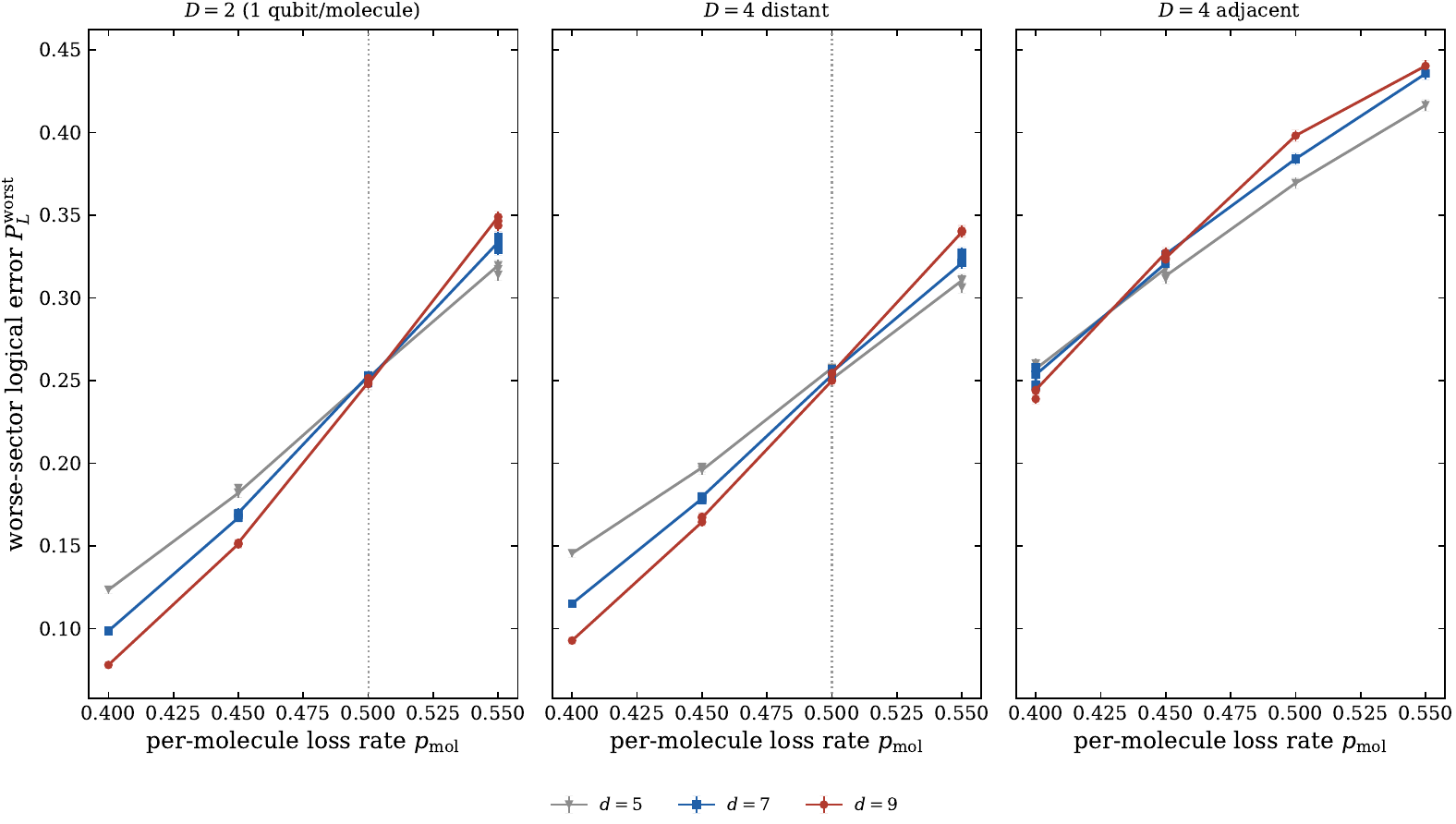}
\caption{Worse sector logical error rate against per molecule loss rate at
$d=5,\;7,\;9$, pooled over three independent runs of $2\times10^{4}$ shots per sector.
The dotted line marks $p_{\mathrm{mol}}=0.50$ and is drawn only in the two panels
where the curves cross there. Error bars are binomial.}
\label{fig:crossing}
\end{figure*}

We begin by asking what the $D=4$ encoding costs relative to $D=2$. In this section, we
use the distant embedding, and we return to the question of embedding geometry in
Sec.~\ref{sec:anisotropy}. Figure~\ref{fig:crossing} shows the worse sector logical error rate against per
molecule loss rate at three code distances. Each point is pooled over three
independent runs of $2\times10^{4}$ shots per sector. The spread between replicates
was at or below the binomial standard error at every point, which is consistent with
there being no systematic run to run effect.
Before comparing the two encodings it is worth noting where the crossing should be
expected. With $f=0$ every molecular loss is heralded, and with perfect projective
stabiliser measurement the channel is pure erasure. Each code qubit is erased exactly
when its molecule is lost, so the per qubit erasure probability equals
$p_{\mathrm{mol}}$ for both encodings, and the relevant reference point is the known
surface code erasure threshold of one half~\cite{stace2010,barrett2010}, which follows
from bond percolation on the square lattice.

For $D=2$ and for $D=4$ with the distant embedding the behaviour is similar. At
$p_{\mathrm{mol}}=0.45$ the curves decrease with $d$; at $0.50$ they coincide to
within error; at $0.55$ they increase with $d$. The crossing therefore falls close to
$p_{\mathrm{mol}}=0.50$ in both cases, and we cannot separate the two crossings at our
statistics. For $D=2$ this recovers the expected erasure value, and we take it as a
check on the implementation and not as a result. For $D=4$ distant the same location
indicates that pairwise correlation between erased qubits, when those qubits are
spatially dispersed, does not noticeably move the crossing. We would emphasise that
these are finite distance crossing estimates and not threshold determinations. Three
distances are not enough to support a finite size scaling fit, and we do not attempt
one. What the data support is the relative statement that the two conditions cross at
the same loss rate to within our resolution, and that this rate is consistent with the
uncorrelated erasure value.

Below the crossing the two encodings differ. At $p_{\mathrm{mol}}=0.45$ and $d=9$ we
obtain $P_L^{\mathrm{worst}}=0.1514(25)$ for $D=2$ and $0.1663(26)$ for $D=4$ distant,
an increase of about $10\%$. The corresponding increases are $8\%$ at $d=5$ and $6\%$
at $d=7$. The difference becomes smaller as the loss rate approaches the crossing,
falling to $1.4\%$ at $d=9$ and $p_{\mathrm{mol}}=0.50$, which is consistent with zero
at our statistics. Above the crossing, at $p_{\mathrm{mol}}=0.55$, the $D=4$ curve
lies slightly below the $D=2$ curve at all three distances, by between $2\%$ and
$3\%$. The sign is the same at every distance, but the individual differences are only
two to four standard errors, so we do not attach significance to this above crossing
behaviour.

The third panel of Fig.~\ref{fig:crossing} shows the same measurement for the adjacent
embedding. Its curves cross at a visibly lower loss rate, below
$p_{\mathrm{mol}}=0.45$, and the separation between distances is smaller throughout
the range we sampled. Unlike the first two panels, this does not sit at the
uncorrelated erasure value, which suggests that the alignment of the correlated loss
units with a lattice direction, and not the presence of correlation as such, is what
moves the crossing. We do not quote a crossing value in this case, since the three
curves nearly coincide over the interval where they meet and an estimate read from the
figure would not be meaningful at that precision. The origin of this behaviour is the
subject of Sec.~\ref{sec:anisotropy}.

Taking the first two panels together, the $D=4$ distant encoding reaches the same
finite distance crossing as $D=2$ while using $48$ to $49\%$ fewer molecules
(Table~\ref{tab:counts}), and pays for this with a sub threshold logical error rate
roughly $6$ to $10\%$ higher over the range of loss rates we sampled. We should be
clear that this range, $p_{\mathrm{mol}}=0.40$ to $0.55$, sits well above the loss
rates at which a code would actually be operated, and was chosen so that the crossing
behaviour is visible at accessible shot counts. Whether the trade is favourable at
realistic loss rates is not something our data can settle, and the deeply sub
threshold regime would be a natural target for future work. We would also note that
our comparison assumes the per molecule loss rate to be the same for both encodings.
If occupying a larger rotational manifold increases the loss rate, the comparison
would shift accordingly.

\section{Embedding induced logical sector anisotropy}
\label{sec:anisotropy}

The comparison above used one particular embedding. We now fix the encoding at $D=4$
and vary only how the two co-located qubits are placed in the code lattice.

\subsection{Sector asymmetry}
Table~\ref{tab:aniso} lists the two sector error rates at $d=9$ and
$p_{\mathrm{mol}}=0.45$. The $D=2$ encoding is symmetric to within error, as it should
be, and we take this as a check on our $X$ sector implementation, which uses a
distinct check matrix, logical operator, and state preparation ($\ket{+}_L$ in place
of $\ket{0}_L$). The distant embedding is also symmetric, with $|P_L^{Z}-P_L^{X}|$
averaging $0.005$ across the grid we sampled and never exceeding $0.014$, which is
comparable to the $D=2$ control at $0.003$ and $0.007$. The adjacent embedding is not.
At this operating point the two sectors differ by $0.27$, roughly fifty times the
distant value.
\begin{table}
\caption{Sector error rates at $d=9$, $p_{\mathrm{mol}}=0.45$ ($2\times10^{4}$ shots
per sector).}
\label{tab:aniso}
\begin{ruledtabular}
\begin{tabular}{lcccc}
 & $P_L^{Z}$ & $P_L^{X}$ & $P_L^{\mathrm{worst}}$ & $N$ \\ \colrule
$D=2$          & 0.151 & 0.151 & 0.151 & 81 \\
$D=4$ adjacent & 0.059 & 0.327 & 0.327 & 45 \\
$D=4$ distant  & 0.164 & 0.164 & 0.164 & 41
\end{tabular}
\end{ruledtabular}
\end{table}
The direction of the asymmetry is the one suggested by Fig.~\ref{fig:schematic}(c).
The adjacent construction produces erasure dimers lying parallel to $\bar{Z}$, and it
is the $X$ sector that suffers. We have not attempted to derive the size of the effect
from the geometry, and we present the alignment as a consistent observation and not as
a demonstrated mechanism.

\subsection{Loss of distance suppression}

\begin{figure*}
\includegraphics[width=0.8\textwidth]{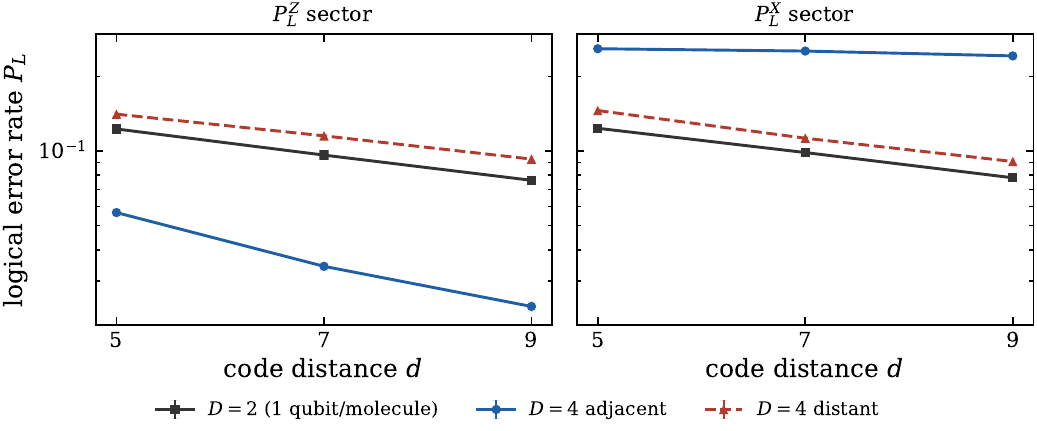}
\caption{Logical error rate against code distance at $p_{\mathrm{mol}}=0.40$,
resolved by logical sector, on a shared logarithmic axis ($2\times10^{4}$ shots per
point). In the $Z$ sector all three architectures suppress error with distance. In
the $X$ sector the $D=4$ adjacent embedding shows little or no suppression over $d=5$
to $9$, while $D=2$ and $D=4$ distant behave as they do in the $Z$ sector.}
\label{fig:aniso}
\end{figure*}

The asymmetry on its own might be read as a prefactor, in principle correctable by
operating at lower loss. The scaling with distance suggests a different reading.
Figure~\ref{fig:aniso} shows the logical error rate against code distance at
$p_{\mathrm{mol}}=0.40$, resolved by sector. In the favoured $Z$ sector the adjacent
embedding suppresses logical error with $d$ in the usual way, by a factor of $2.3$
between $d=5$ and $d=9$ at $p_{\mathrm{mol}}=0.40$ and a factor of $1.5$ at $0.45$. In
the disfavoured $X$ sector, over the same range, we see little or no suppression. At
$p_{\mathrm{mol}}=0.40$ the values are $0.2557(31)$, $0.2577(31)$ and $0.2389(30)$ for
$d=5,7,9$, which is non monotonic and amounts to a net change of $7\%$ over two
distance steps. At $p_{\mathrm{mol}}=0.45$ they are $0.3177(33)$, $0.3209(33)$ and
$0.3272(33)$, so mildly increasing with $d$, though the $d=5$ to $d=9$ difference is
only two standard errors.
We interpret this result cautiously. Three distances cannot establish asymptotic behaviour, and we cannot exclude slow suppression that our range does not resolve. What we can say is that over \(d=5\) to \(9\) the disfavoured sector of the adjacent embedding shows little benefit from increasing code distance, while the favoured sector of the same code, at the same loss rate and shot count, improves by a factor of \(1.5\) to \(2.3\). We would also note that the \(D=2\) and distant curves have visibly similar distance dependence in both panels of Fig.~\ref{fig:aniso}, which offers some further reassurance that the \(X\) sector implementation is sound and that the anomaly is confined to the adjacent case. If the behaviour we observe persists, the disfavoured sector would be at or above its crossing point at loss rates where \(D=2\) and \(D=4\) distant remain comfortably below theirs, and the deficit would not be recovered by building a larger code. This is consistent with the third panel of Fig.~\ref{fig:crossing}, where the adjacent crossing falls below \(p_{\mathrm{mol}}=0.45\).
\subsection{Implication for benchmarking}

The practical lesson we would draw is a methodological one. Had we characterised the
adjacent embedding in the $Z$ sector alone, we would have reported a logical error
rate roughly three times better than $D=2$ at the same loss rate, together with a
$44\%$ reduction in molecule count, which would have looked like a strong result. The
$X$ sector reverses that conclusion. For architectures in which correlated loss has a
directional geometry, we would suggest that single sector benchmarks are not by
themselves sufficient evidence, and that a worse sector figure of merit is the safer
default.

More generally, the two $D=4$ cases differ only in the embedding, yet they differ in
worse sector error by a factor of two and differ qualitatively in scaling behaviour.
The molecular encoding dimension alone therefore does not determine logical
performance. The spatial embedding of the correlated loss units is an additional
design parameter, and in our simulations it dominates the difference between the two
$D=4$ layouts.

\section{Imperfect heralding}
\label{sec:heralding}

The results above assume that every molecular loss is heralded, so that the decoder
receives the location of each erased qubit. This is optimistic. Population transferred
to internal states outside the readout manifold is lost without a detection event, and
which leakage channels are heraldable is set by the rotational structure of the
molecule. We therefore repeated the comparison with a fraction $f$ of losses made
silent, so that the affected qubits receive a random Pauli but the decoder is given no
zero weight edge for them. We include the $D=2$ encoding as a reference here, since
the question of interest is whether the molecule saving of the $D=4$ encoding survives
imperfect heralding.

\begin{figure}
\includegraphics[width=0.9\columnwidth]{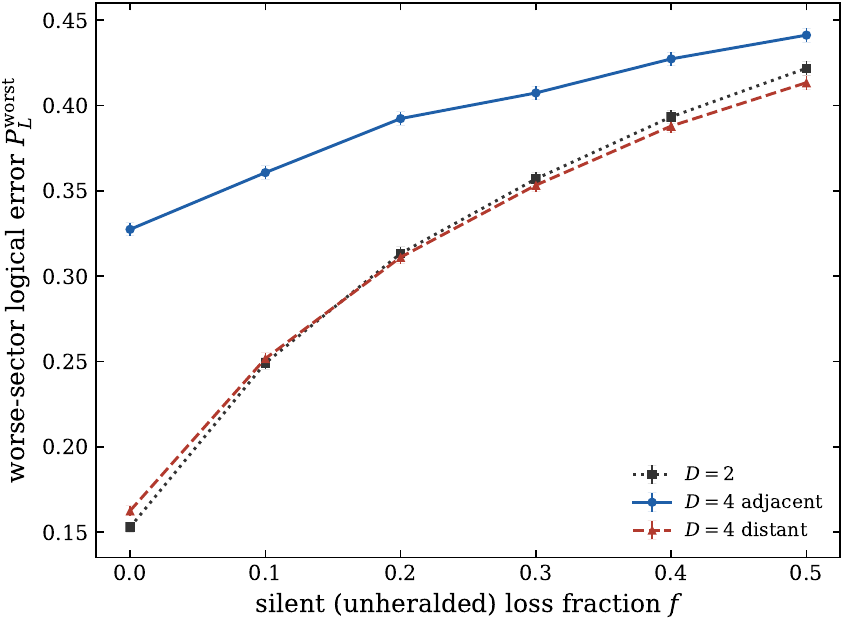}\\[2ex]
\includegraphics[width=0.9\columnwidth]{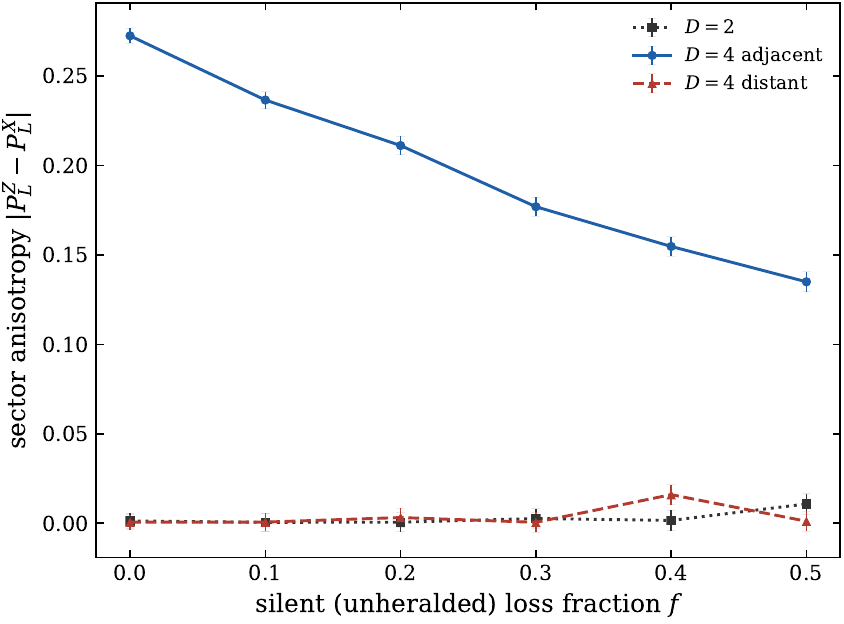}
\caption{Effect of unheralded loss at $d=9$, $p_{\mathrm{mol}}=0.45$
($1.5\times10^{4}$ shots per sector, single run per point). Top: worse sector logical
error against silent fraction $f$. Bottom: sector anisotropy $|P_L^{Z}-P_L^{X}|$
against $f$. In the lower panel the $D=2$ and $D=4$ distant curves overlap near zero.}
\label{fig:herald}
\end{figure}

All three architectures degrade with $f$, as they must, since at $f=0.5$ half of all
erasures reach the decoder as undiagnosed Pauli errors. The worse sector error at
$d=9$ and $p_{\mathrm{mol}}=0.45$ rises from $0.153$ to $0.422$ for $D=2$, from
$0.163$ to $0.413$ for $D=4$ distant, and from $0.327$ to $0.441$ for $D=4$ adjacent.

The comparison that matters for the encoding choice is the ratio to the $D=2$
baseline, given in Table~\ref{tab:herald}. At perfect heralding the $D=4$ distant
embedding carries the $6\%$ penalty reported in Sec.~\ref{sec:encoding}. That penalty
becomes smaller as $f$ increases and is gone by $f\approx0.2$; beyond that the ratio
sits slightly below unity, between $0.98$ and $0.99$. We would not read the apparent
reversal as real, since the differences are under $1\%$ against a single run standard
error of about $0.004$ per point and no individual ratio below unity is significant.
The disappearance of the penalty, on the other hand, is consistent across four
successive values of $f$ and seems to us reasonably robust.

\begin{table}
\caption{Worse sector logical error at $d=9$, $p_{\mathrm{mol}}=0.45$, and its ratio
to the $D=2$ baseline, as a function of the unheralded fraction $f$. Single run of
$1.5\times10^{4}$ shots per sector per point.}
\label{tab:herald}
\begin{ruledtabular}
\begin{tabular}{ccccc}
$f$ & $D=2$ & $D=4$ dist. & $D=4$ adj. & dist./$D{=}2$ \\ \colrule
0.0 & 0.153 & 0.163 & 0.327 & 1.06 \\
0.1 & 0.249 & 0.252 & 0.361 & 1.01 \\
0.2 & 0.313 & 0.311 & 0.392 & 0.99 \\
0.3 & 0.357 & 0.353 & 0.407 & 0.99 \\
0.4 & 0.393 & 0.388 & 0.427 & 0.99 \\
0.5 & 0.422 & 0.413 & 0.441 & 0.98
\end{tabular}
\end{ruledtabular}
\end{table}

The reading we would offer is that imperfect heralding does not penalise the $D=4$
encoding relative to $D=2$. If anything the encoding trade improves, since the modest
sub threshold cost of the molecule saving is a feature of the well heralded limit and
is paid for by the same location information that $D=2$ also loses when heralding
fails. An architecture with poor heralding would therefore have less reason, and not
more, to avoid the $D=4$ encoding on error rate grounds.

The two $D=4$ embeddings behave differently from one another. The gap between them
narrows from a factor of $2.0$ at $f=0$ to $1.07$ at $f=0.5$, which we would attribute
to the adjacent embedding's disfavoured sector already performing poorly at $f=0$ and
having less left to lose. We would caution against reading this convergence as the two
embeddings becoming equivalent. The sector structure, shown in the lower panel of
Fig.~\ref{fig:herald}, is largely unaffected by heralding. The anisotropy of the
distant embedding stays between $0.001$ and $0.016$ with no trend, indistinguishable
from the $D=2$ control over the same range, while the adjacent embedding remains
anisotropic at every $f$, declining smoothly from $0.273$ to $0.135$ as unheralded
loss comes to dominate both of its sectors. The adjacent embedding is still the one
with a sector its decoder cannot protect, and unheralded loss merely raises the floor
under the other sector until the difference is less visible.

Within this slice, then, the two effects appear to be approximately separable. Heralding quality
sets the overall magnitude of the logical error and the size of the gap between
embeddings, while the embedding geometry sets the sector asymmetry and does so
independently of $f$. We would note the limits of the statement. It is a single
distance and a single loss rate, sampled at six values of $f$ with one run each, and
it establishes the direction of the $f$ dependence and not its functional form. In
particular we have not determined whether the finite distance crossing point itself
moves with $f$, which is what would decide how much heralding efficiency a given
architecture requires.

\section{Coherent dipolar exchange}
\label{sec:coherent}
The exchange interaction of Eq.~\eqref{eq:hex} is present continuously, including
during idling and syndrome extraction, and is not switched off between gates. A
decoder built on a Pauli noise model does not represent it directly, and the usual
approximation is to replace the coherent evolution by its Pauli twirl. We therefore
asked how much that approximation costs, and in which direction it errs.

\subsection{Comparison with the Pauli twirl}

\begin{figure}
\includegraphics[width=0.9\columnwidth]{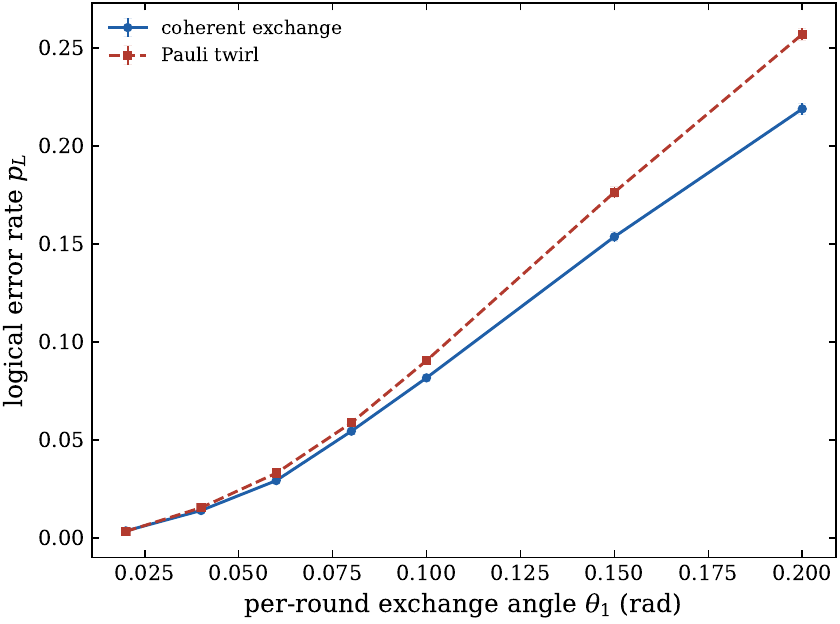}\\[2ex]
\includegraphics[width=0.9\columnwidth]{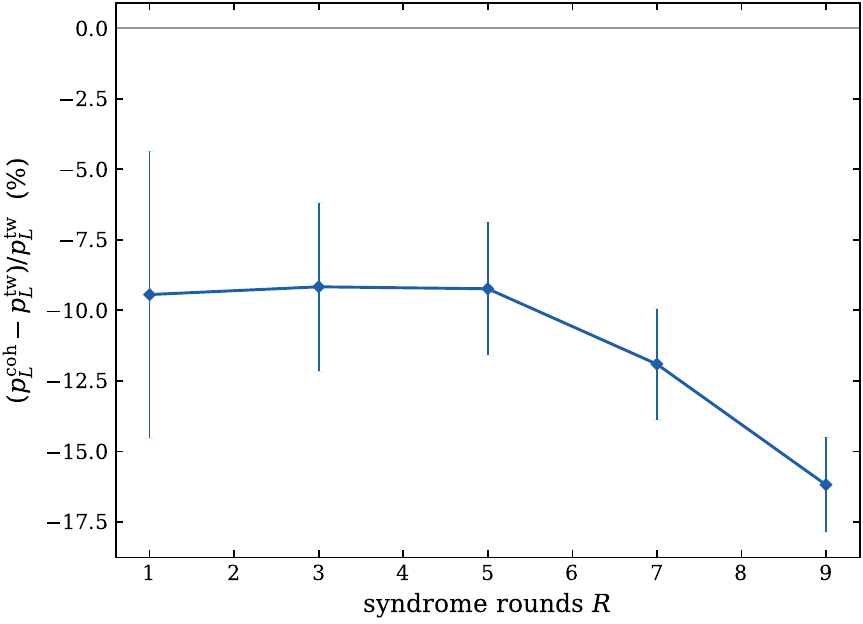}
\caption{Coherent exchange compared with its Pauli twirl at $d=3$, $2\times10^{4}$
shots per point. Top: logical error against per round nearest neighbour exchange angle
$\theta$. Bottom: relative discrepancy
$(P_L^{\mathrm{coh}}-P_L^{\mathrm{tw}})/P_L^{\mathrm{tw}}$ against the number of
syndrome rounds $R$ at $\theta=0.10$.}
\label{fig:coherent}
\end{figure}

Figure~\ref{fig:coherent} compares the two arms at $d=3$ with $2\times10^{4}$ shots
per point. Across the range $\theta=0.04$ to $0.20$ the coherent evolution gives a
lower logical error rate than its twirl, by $7$ to $15\%$. The discrepancy grows with
$\theta$ and reaches $-14.8\%$ at $\theta=0.20$, where it is nine standard errors. At
the smallest angle we sampled, $\theta=0.02$, the two arms are indistinguishable,
which is expected once the logical error rate approaches the shot noise floor.
The direction of this result is opposite to the one familiar from single qubit
coherent over rotations, for which the twirl is generally
optimistic~\cite{bravyi2018}. It is, however, consistent with reports that the Pauli
twirl approximation can be pessimistic for other noise
models~\cite{tomita2014,gutierrez2015}, and we would regard our observation as a
further instance and not a new phenomenon.

We also varied the number of syndrome rounds $R$ at fixed $\theta=0.10$, decoupling
$R$ from the code distance, in order to test whether the discrepancy is eroded by
repeated measurement. It does not appear to be. The relative gap is $-9.4\%$ at $R=1$
and grows in magnitude to $-16.2\%$ at $R=9$. Whatever the twirl discards is therefore
not removed by measurement back action over the range we tested. Both logical sectors show the same sign. At $\theta=0.12$ with $4\times10^{3}$ shots
we find $-16.4\%$ in the $Z$ sector and $-5.7\%$ in the $X$ sector. The magnitudes
differ, which is expected, since the generator $(X_iX_j+Y_iY_j)/2$ is symmetric under
$X\leftrightarrow Y$ but not under $X\leftrightarrow Z$ and there is no reason for it
to act identically on the two sectors.

\subsection{Origin of the discrepancy}

To locate the effect we instrumented the syndrome distribution at $d=3$ and
$\theta=0.12$. The coherent arm produces more defects on average ($0.409$ against
$0.324$) and heavier matchings ($0.368$ against $0.300$), so the advantage does not
come from generating fewer or cleaner errors. The conditional failure probability at
fixed syndrome weight is also nearly identical between the two arms ($0.064$ against
$0.070$ at weight two), so it does not come from the decoder resolving a given
syndrome better.

\begin{figure}
\includegraphics[width=0.9\columnwidth]{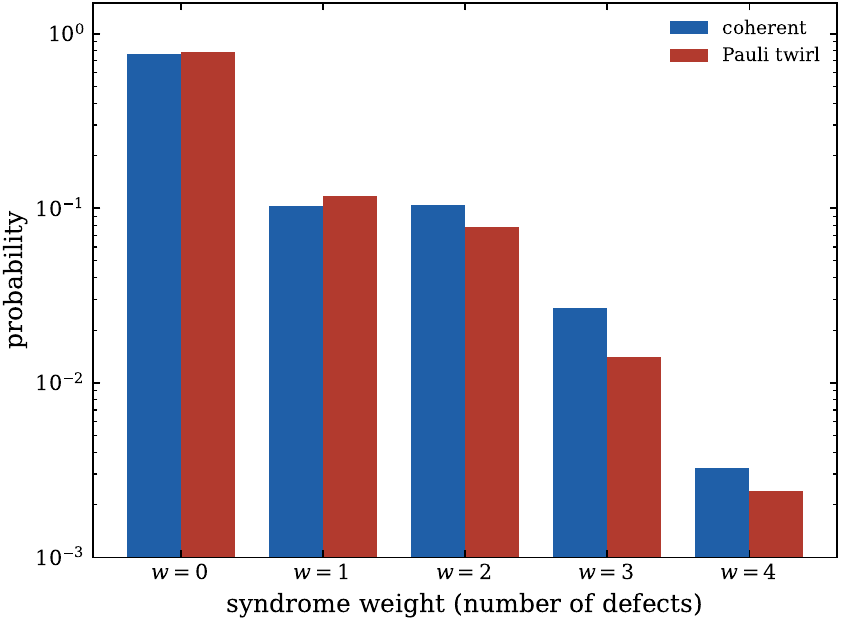}
\caption{Syndrome weight distribution at $d=3$, $\theta=0.12$. The coherent arm places
less probability on weight one, which fails with certainty at this distance, and more
on weights two and higher, which fail with probability $0.07$ and below. Conditional
failure probabilities are $P_{\mathrm{fail}}=1.00$ at $w=1$, $0.07$ at $w=2$, and $0$
at $w=3,4$.}
\label{fig:mechanism}
\end{figure}

What differs is the distribution of syndrome weights, shown in
Fig.~\ref{fig:mechanism}. The exchange interaction conserves total excitation number
and so moves defects in pairs, and the coherent arm places less probability on odd
weight syndromes and more on paired ones. At $d=3$ a single isolated defect matches to
a boundary and flips the logical operator with certainty, so this redistribution
matters. The weight one probability is $0.100$ coherent against $0.117$ twirled, with
the difference appearing at weights two and three, which fail with probability $0.06$
and $0$ respectively. Applying the twirled arm's conditional failure probabilities to
the coherent arm's syndrome distribution reproduces the coherent logical error rate
($0.1079$ against $0.1072$ measured, with $0.1220$ for the twirl), which indicates
that the syndrome distribution accounts for essentially the whole effect.
We would not want to extend this account too far. The fatal weight one channel is a
feature of $d=3$, where a single defect necessarily reaches a boundary, and it need
not remain the dominant failure mode at larger distances. The exponent of $P_L$ in
$\theta$ is the same for both arms to within our resolution, which suggests a
prefactor effect and not a change in effective distance. Establishing whether the
discrepancy survives at larger $d$ would require tensor network or other approximate
methods, since exact state vector simulation is limited to $d=3$, and we have not
attempted this. Our data show that, in the regime accessible to exact simulation, the Pauli channel approximation overestimates the logical error rate of the exchange interaction. The excitation conserving structure of the interaction provides a plausible explanation for this discrepancy.

\section{Discussions and Future Outlook}
\label{sec:discussion}

Taken together, our results suggest three considerations for surface code error
correction on molecular tweezer arrays. The $D=4$ encoding appears to be a usable hardware saving. It reaches the same finite
distance crossing point as $D=2$ with roughly half the molecules, at a sub-threshold
logical error cost of $6$-$10\%$ in the regime of interest. Whether this is a good
trade depends on the relative difficulty of adding molecules and of tolerating logical
error in a particular experiment.

The saving is contingent on the embedding. Placing the two co-located qubits along a
lattice direction produces correlated erasure clusters aligned with a logical string
and, in our simulations, a logical sector asymmetry of roughly fifty times, with the
disfavoured sector showing little improvement with code distance over $d=5$ to $9$.
Since the encoding is identical in the two cases, the embedding appears to be a design parameter in its own right. Our results suggest avoiding embeddings in which the correlated loss units are aligned with a logical operator. The two constructions considered here represent two distinct points in the broader space of possible embeddings, which remains to be explored. Heralding efficiency sets the size of the benefit but not the sector structure. Within
the single slice we examined, unheralded loss degrades the dispersed embedding faster
than the aligned one and narrows their separation, while leaving the sector anisotropy
of each essentially unchanged, and it does not penalise the $D=4$ encoding relative to
$D=2$. Since which molecular leakage channels are heraldable is determined by the
rotational level structure, this connects a decoder level quantity to a molecular
structure one. As some possible extension, it is worth to find whether the finite distance
crossing point itself depends on the unheralded fraction $f$. In other words, whether
imperfect heralding costs logical error rate or heralding efficiency becomes a requirement on the molecular level structure.
One may also extend the coherent exchange comparison beyond $d=3$ using tensor network methods.

\section*{Data and code availability}

The numerical data underlying the figures are provided in Appendix A. Additional numerical data will be made available upon reasonable request to the authors.

\section*{Acknowledgement}
NV and UR acknowledge the financial support from Indian National Quantum Mission (Quantum Computing).


\bibliography{reference}

\clearpage
\onecolumngrid
\appendix

\section{Tabulated data}
\label{app:data}

The numerical values underlying the figures are collected below. Additional numerical data may be made available from the author upon reasonable request. Table~\ref{tab:a1} gives the sector resolved logical error rates under heralded
molecular loss, from which Figs.~\ref{fig:crossing} and~\ref{fig:aniso} are drawn.
Values are pooled over replicates where more than one run was performed. The $D=2$
rows serve as the sector symmetry control discussed in Sec.~\ref{sec:anisotropy}. Table~\ref{tab:a2} gives the coherent and Pauli twirled logical error rates at $d=3$
for both the sweep in exchange angle and the sweep in the number of syndrome rounds,
underlying Fig.~\ref{fig:coherent}. Table~\ref{tab:a3} gives the syndrome weight distribution behind
Fig.~\ref{fig:mechanism}, together with the conditional failure probabilities used in
the counterfactual calculation of Sec.~\ref{sec:coherent}.
\begin{table}[!b]
\caption{Sector resolved logical error rates under heralded molecular loss ($f=0$). Values are pooled over all available replicates; parenthesised figures are binomial standard errors in the last two digits.}
\label{tab:a1}
\begin{ruledtabular}
\begin{tabular}{llccc}
encoding & $d$ & $p_{\mathrm{mol}}$ & $P_L^{Z}$ & $P_L^{X}$ \\
\colrule
$D{=}2$ & 5 & 0.40 & 0.1226(23) & 0.1235(23) \\
 & 5 & 0.45 & 0.1829(16) & 0.1811(16) \\
 & 5 & 0.50 & 0.2478(18) & 0.2498(18) \\
 & 5 & 0.55 & 0.3160(19) & 0.3138(19) \\
 & 7 & 0.40 & 0.0963(21) & 0.0986(21) \\
 & 7 & 0.45 & 0.1670(15) & 0.1677(15) \\
 & 7 & 0.50 & 0.2471(18) & 0.2508(18) \\
 & 7 & 0.55 & 0.3330(19) & 0.3308(19) \\
 & 9 & 0.40 & 0.0762(19) & 0.0780(19) \\
 & 9 & 0.45 & 0.1512(15) & 0.1512(15) \\
 & 9 & 0.50 & 0.2484(18) & 0.2482(18) \\
 & 9 & 0.55 & 0.3445(19) & 0.3441(19) \\
\colrule
$D{=}4$ dist. & 5 & 0.40 & 0.1407(25) & 0.1455(25) \\
 & 5 & 0.45 & 0.1903(16) & 0.1969(16) \\
 & 5 & 0.50 & 0.2479(18) & 0.2535(18) \\
 & 5 & 0.55 & 0.3024(19) & 0.3086(19) \\
 & 7 & 0.40 & 0.1151(23) & 0.1126(22) \\
 & 7 & 0.45 & 0.1766(16) & 0.1787(16) \\
 & 7 & 0.50 & 0.2510(18) & 0.2542(18) \\
 & 7 & 0.55 & 0.3198(19) & 0.3234(19) \\
 & 9 & 0.40 & 0.0928(21) & 0.0907(20) \\
 & 9 & 0.45 & 0.1647(15) & 0.1641(15) \\
 & 9 & 0.50 & 0.2483(18) & 0.2515(18) \\
 & 9 & 0.55 & 0.3368(19) & 0.3394(19) \\
\colrule
$D{=}4$ adj. & 5 & 0.40 & 0.0564(9) & 0.2576(18) \\
 & 5 & 0.45 & 0.0887(12) & 0.3142(19) \\
 & 5 & 0.50 & 0.1288(24) & 0.3694(34) \\
 & 5 & 0.55 & 0.1810(27) & 0.4164(35) \\
 & 7 & 0.40 & 0.0347(7) & 0.2528(18) \\
 & 7 & 0.45 & 0.0679(10) & 0.3241(19) \\
 & 7 & 0.50 & 0.1155(23) & 0.3840(34) \\
 & 7 & 0.55 & 0.1752(27) & 0.4355(35) \\
 & 9 & 0.40 & 0.0245(6) & 0.2423(17) \\
 & 9 & 0.45 & 0.0556(9) & 0.3248(19) \\
 & 9 & 0.50 & 0.1009(21) & 0.3981(35) \\
 & 9 & 0.55 & 0.1714(27) & 0.4402(35) \\
\end{tabular}
\end{ruledtabular}
\end{table}

\begin{table}[!h]
\caption{Coherent exchange compared with its Pauli twirl at $d=3$. Upper block: sweep in the per round exchange angle $\theta$ at $R=d=3$ rounds. Lower block: sweep in the number of syndrome rounds $R$ at fixed $\theta=0.10$. The final column is the relative discrepancy $(P_L^{\mathrm{coh}}-P_L^{\mathrm{tw}})/P_L^{\mathrm{tw}}$.}
\label{tab:a2}
\begin{ruledtabular}
\begin{tabular}{lccc}
 & $P_L^{\mathrm{coh}}$ & $P_L^{\mathrm{tw}}$ & rel. (\%) \\
\colrule
$\theta=0.02$ & 0.0036(4) & 0.0034(4) & $+7.5$ \\
$\theta=0.04$ & 0.0140(8) & 0.0155(9) & $-9.4$ \\
$\theta=0.06$ & 0.0293(12) & 0.0331(13) & $-11.5$ \\
$\theta=0.08$ & 0.0544(16) & 0.0587(17) & $-7.3$ \\
$\theta=0.10$ & 0.0817(19) & 0.0905(20) & $-9.7$ \\
$\theta=0.15$ & 0.1536(25) & 0.1763(27) & $-12.9$ \\
$\theta=0.20$ & 0.2188(29) & 0.2569(31) & $-14.8$ \\
\colrule
$R=1$ & 0.0293(12) & 0.0323(13) & $-9.4$ \\
$R=3$ & 0.0808(19) & 0.0889(20) & $-9.2$ \\
$R=5$ & 0.1238(23) & 0.1364(24) & $-9.2$ \\
$R=7$ & 0.1572(26) & 0.1784(27) & $-11.9$ \\
$R=9$ & 0.1830(27) & 0.2184(29) & $-16.2$ \\
\end{tabular}
\end{ruledtabular}
\end{table}

\begin{table}
\caption{Syndrome weight distribution at $d=3$, $\theta=0.12$, and the conditional
failure probability at each weight. The final row applies the twirled conditional
failure probabilities to the coherent weight distribution.}
\label{tab:a3}
\begin{ruledtabular}
\begin{tabular}{lccc}
$w$ & $P_{\mathrm{coh}}(w)$ & $P_{\mathrm{tw}}(w)$ & $P_{\mathrm{fail}}(w)$ \\
\colrule
0 & 0.762 & 0.789 & 0.00 \\
1 & 0.100 & 0.117 & 1.00 \\
2 & 0.108 & 0.078 & 0.07 \\
3 & 0.026 & 0.014 & 0.00 \\
4 & 0.004 & 0.003 & 0.00 \\
\colrule
\multicolumn{4}{l}{measured $P_L$: coherent $0.1072$, twirl $0.1220$} \\
\multicolumn{4}{l}{coherent $P(w)$ with twirled $P_{\mathrm{fail}}(w)$: $0.1079$} \\
\end{tabular}
\end{ruledtabular}
\end{table}
\end{document}